\documentclass[aps,pra,reprint,groupedaddress]{revtex4-1}
\usepackage{graphicx}
\usepackage{braket}
\usepackage{amssymb}
\usepackage{hyperref}
\usepackage{physics}
\hypersetup{
    colorlinks=true,
    linkcolor=blue,
    citecolor=blue,
    urlcolor=blue,
}

\usepackage[overload]{empheq}
\newcommand{\for}{\text{for }}

\usepackage{amsmath}

\begin{document}


\title{High-dimensional quantum teleportation under noisy environments}


\author{Alejandro Fonseca}
\affiliation{Departamento de F\'{\i}sica, CCEN,  Universidade Federal 
de Pernambuco, 50670-901, Recife, PE, Brazil}


\date{\today}

\begin{abstract}
We study the protocol of qudit teleportation using quantum systems subjected to several kinds of noise for arbitrary dimensionality $d$. We consider four classes of noise: dit-flip, $d$-phase-flip, dit-phase-flip and depolarizing, each of them corresponding to a family of Weyl operators, introduced via Kraus formalism. We derive a general expression for the average fidelity of teleportation in arbitrary dimension $d$ for any combination of noise on the involved qudits. Under a different approach we derive the average fidelity of teleportation for a more general scenario involving the $d$-dimensional generalization of amplitude damping noise as well. We show that all possible scenarios may be classified in four different behaviours and discuss the cases in which it is possible to improve the fidelity by increasing the associated noise fractions. All our results are in agreement with previous analysis by Fortes and Rigolin for the case of qubits (\hyperlink{https://doi.org/10.1103/PhysRevA.92.012338}{Phys. Rev. A, 92 012338, 2015}).
\end{abstract}


\maketitle

\section{Introduction}

Since its proposal in 1993 \cite{Bennett93} until nowadays, the teleportation protocol represents one of the most known and widely studied applications of quantum entanglement \cite{Pirandola15}. According to it, if one has a source of maximally entangled qudits and a measurement apparatus capable of discriminating the $d^2$ elements of the generalized Bell basis, then it is possible to send an arbitrary qudit state between two locations even without prior knowledge of it. Nevertheless, in day by day experiments some additional features arise from the unavoidable interaction of the involved parts with the environment and/or imperfections in the preparation of the system, leading to losses of the resources responsible for the improvement in the execution of the task over its classical analogue. Thus, in addition to adopting strategies aiming to diminish the action of noise, one may also modify the scheme of measurements and operations on the parts and consequently to improve the performance of the protocol \cite{Banaszek00,Taketani12}. For this reason, in order to effectively optimize the strategies to be implemented it is important to have a general picture of the scenario. In this respect, the problem of characterizing the protocol of teleportation in the presence of noisy environments have been addressed from several perspectives: In \cite{Oh2002}, Oh and collaborators employ the Lindblad operators formalism, obtaining the fidelity of teleportation for several classes of noise in the quantum channel. More recently, Fortes and Rigolin have presented a set of results within the frame of Kraus operators for some of the most known instances of noise in the literature \cite{Fortes15}. By using the same approach, the authors of \cite{Knoll14} contrast theoretical predictions with experimental results. In addition, there are even some approaches to the problem of multipartite noisy teleportation \cite{Hu10,Cunha2017,Moreno2018}.


Apart from a few exceptions, most of the effort in describing quantum information protocols under noisy scenarios has been focused on systems involving qubits, nevertheless it has been shown that the performance of several tasks is enhanced when high dimensional systems are used instead \cite{Cerf02,Durt04,Vertesi10,Islam2017,Skrzypczyk2018a}. 
In fact, two independent groups have been able to successfully implement the teleportation protocol of qutrits recently \cite{Hu2019,Luo2019}.
Furthermore, because of the limited efficiency in usual detectors, it is more convenient the usage of single qudit systems in the process of generation and manipulation of high dimensional entanglement, rather than multi-qubit based structures \cite{Kues17}. In this way, the exploration of quantum information protocols employing qudits under realistic conditions becomes an important aspect of quantum information science.

In this paper we present a characterization of the standard protocol of qudit teleportation considering several sources of imperfection, including non-maximally entangled channel and/or joint measurements and qudits susceptible to modifications of their states due to the interaction with their environment.
The paper is organized as follows: first we present the protocol, derive general expressions for the fidelity of teleportation and study the noiseless case. In section \ref{KO} we introduce the Kraus operators used throughout this work to model the effect of noise in the system. Then we consider the simplest case in which noise acts on a single qudit and extend results to more general scenarios involving more than one part affected. The last part is devoted to discuss our main results and present some conclusions.

\section{Teleportation Protocol}

\begin{figure}
	\centering
	\includegraphics[scale=0.55]{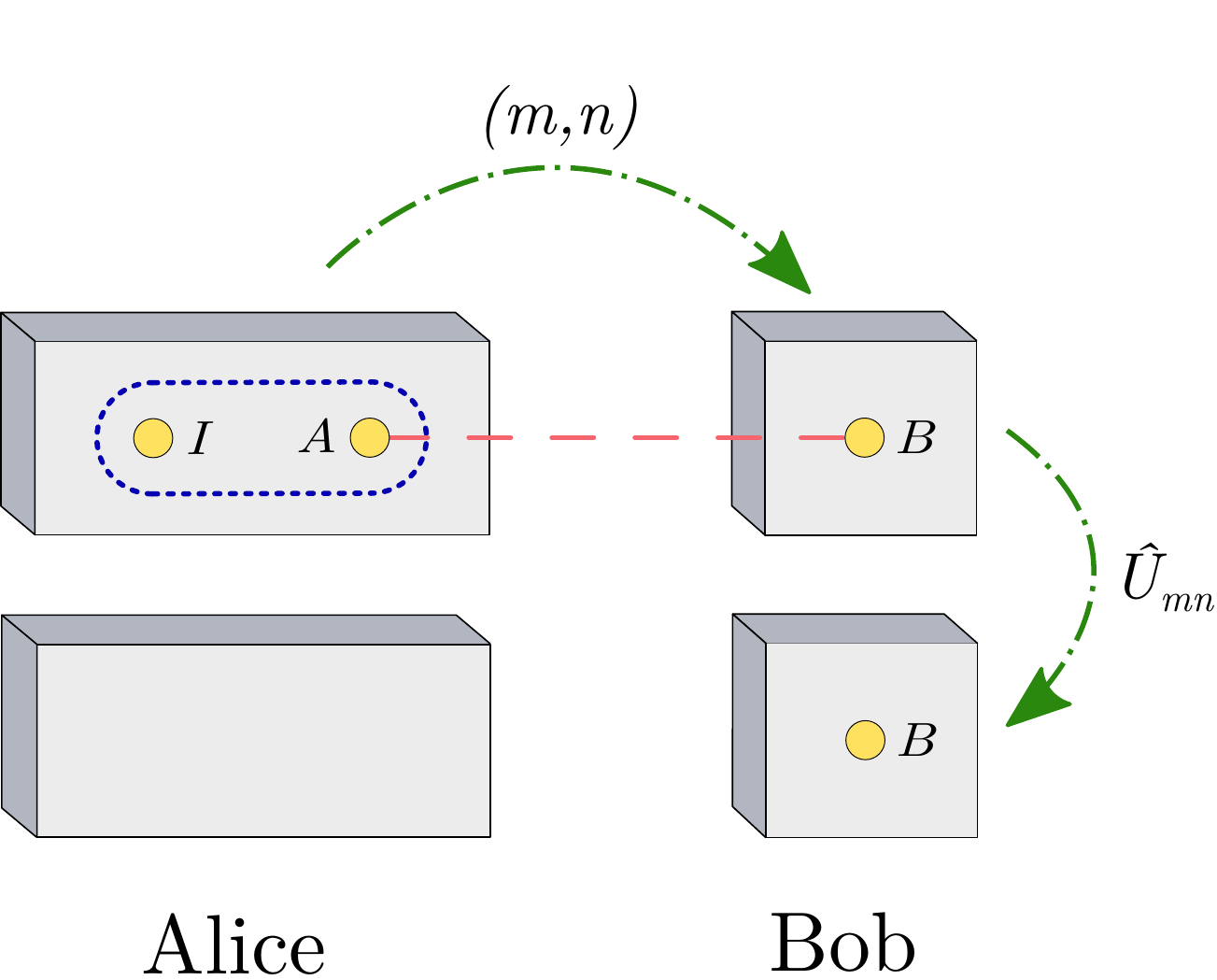}
	\caption{Teleportation scheme: Alice and Bob share a channel composed by two entangled qudits prepared in a state $\hat{\rho}_{ch}$ (red dashed). Alice performs a joint measurement on qudits $I$ and $A$ (blue dotted). Under appropriate conditions (maximal entanglement) and after LOCC (green dot dashed), corresponding to transmission of a pair of dits $(m,n)$ using a classical channel and subsequent application of a local unitary operation $\hat{U}_{mn}$ on qudit $B$, it finally holds in the initial state of the input qudit $I$.}
	\label{TelepS}
\end{figure}

The standard teleportation protocol involves two parts: Alice and Bob, as usual (see Fig. \ref{TelepS}), sharing a pair of entangled qudits whose state may be described by a density operator $\hat{\rho}_{ch}$. Alice has an additional qudit prepared in an arbitrary, not necessarily known qudit state $\ket{\phi}=\sum_{j=0}^{d-1}\alpha_j\ket{j}$, her task is to send it to Bob. For this, she carries out a $d^2$-outcome joint projective measurement on her pair of qudits in a generalized Bell-like basis $\{\ket{\Phi_{mn}}\}$, with elements given by
\begin{equation}
\label{Mb}
\ket{\Phi_{mn}}=\sum_{k=0}^{d-1}\beta_{km}\ket{k,k\oplus n},
\end{equation}
where the symbol ``$\oplus$" denotes sum modulo $d$, the $\beta_{km}$ coefficients account for the extent of entanglement and satisfy the relation $\sum_{k=0}^{d-1}\beta_{km}\beta_{km'}^*=\delta_{mm'}$. In this way, the usual maximally entangled joint measurements are recovered whenever $\beta_{km}=\omega_d^{k\cdot m}/\sqrt{d}$, where $\omega_d=\exp(2\pi i/d)$ is the primitive $d$-th root of unity.

By employing a classical channel, Alice sends the information about her measurement outcome ($m,n$) to Bob who applies a local unitary operation on his qudit, given by one out of the $d^2$ Weyl operators $\hat{U}_{mn}$, defined as \cite{Bertlmann08}
\begin{equation}
\hat{U}_{mn}=\sum_{j=0}^{d-1}\omega_d^{j m}\dyad{j}{j\oplus n}.
\label{UKOp}
\end{equation}

After each run, the state of Bob's qudit (up to normalization) holds
\begin{equation}
\nonumber
\hat{\rho}_{mn}= \hat{U}_{mn}~\tr_{A}\Big\{\Big(\dyad{\Phi_{mn}}{\Phi_{mn}}\otimes \hat{1}_B\Big)\dyad{\phi}{\phi}\otimes\hat{\rho}_{ch} \Big\}~\hat{U}_{mn}^{\dagger},
\end{equation} 
where $\tr_{A}$ denotes the partial trace on the subsystem associated to Alice.

The reliability of the protocol is usually assesed by calculating the fidelity of teleportation i.e. how close the state in Bob is after the process to that initially possessed by Alice: $F_{mn}=\tr \left(\hat{\rho}_{mn} \dyad{\phi}{\phi} \right)\tr \left(\hat{\rho}_{mn}\right)^{-1}$, where the normalization factor $\tr \left(\hat{\rho}_{mn}\right)^{-1}$ is equal to the probability of occurrence of the $(m,n)$ output. By considering the whole possible measurement results, the mean fidelity reads
\begin{equation}
\label{F0}
F=\sum_{mn}\tr\left\{\dyad{\phi}{\phi}\hat{\rho}_{mn}\right\}.
\end{equation} 

Given that our goal is to assess the quality of the protocol, independent of the teleported state and due to the fact that $F$ typically depends on the coefficients $\alpha_j$, then it is more convenient to calculate the average fidelity over the set of input states
\begin{equation}
\label{F2}
\expval{F}=\frac{1}{V_d}\int \dd \Gamma_d ~ F.
\end{equation}
General expressions for $\dd\Gamma_d$ and $V_d$ are given in appendix \ref{ApA}. For the sake of simplicity in what follows we will refer to $\expval{F}$ as the fidelity instead of average fidelity of teleportation.

\subsection{Noise-free environment}
\label{NFE}
Let us assume the qudits composing the channel are initially prepared in a pure partially entangled state, i.e. $\hat{\rho}_{ch}=\dyad{\psi}$, with  $\ket{\psi}=\sum_{k=0}^{d-1}\gamma_k \ket{kk}$. After some calculations the fidelity 
of teleportation is reduced to
\begin{equation*}
\label{FNL}
\expval{F}=f_C\left\{1+ \frac{1}{d}\sum_{\substack{m,n=0\\j>k=0}}^{d-1} \Re\left[\omega_d^{m(k-j)}\beta_{jm}\beta_{km}^*\gamma_{k\oplus n}\gamma_{j\oplus n}^*\right]\right\},
\end{equation*}
where $f_C=\frac{2}{d+1}$ is known as \textit{classical fidelity} and corresponds to the maximal value attained by the fidelity of teleportation after the usage of any classical strategy \cite{Horodecki1999,Weinar13}, e.g. by using a classical channel, Alice sends to Bob the result of a projective measurement carried out on the input qudit. Then, based on this information he guesses the state his qudit has to be prepared.

Remarkably, in the expression above the classical and quantum contributions to the fidelity are made clear. Furthermore, it is easy to see that whenever we have maximal entanglement in the measurements and channel $(\gamma_k=1/\sqrt{d})$, the second term is reduced to $(d-1)/(d+1)$, and in this way the fidelity reaches its maximum value, $\expval{F}=1$, as expected.

\begin{figure}
	\centering
	\includegraphics[scale=0.36]{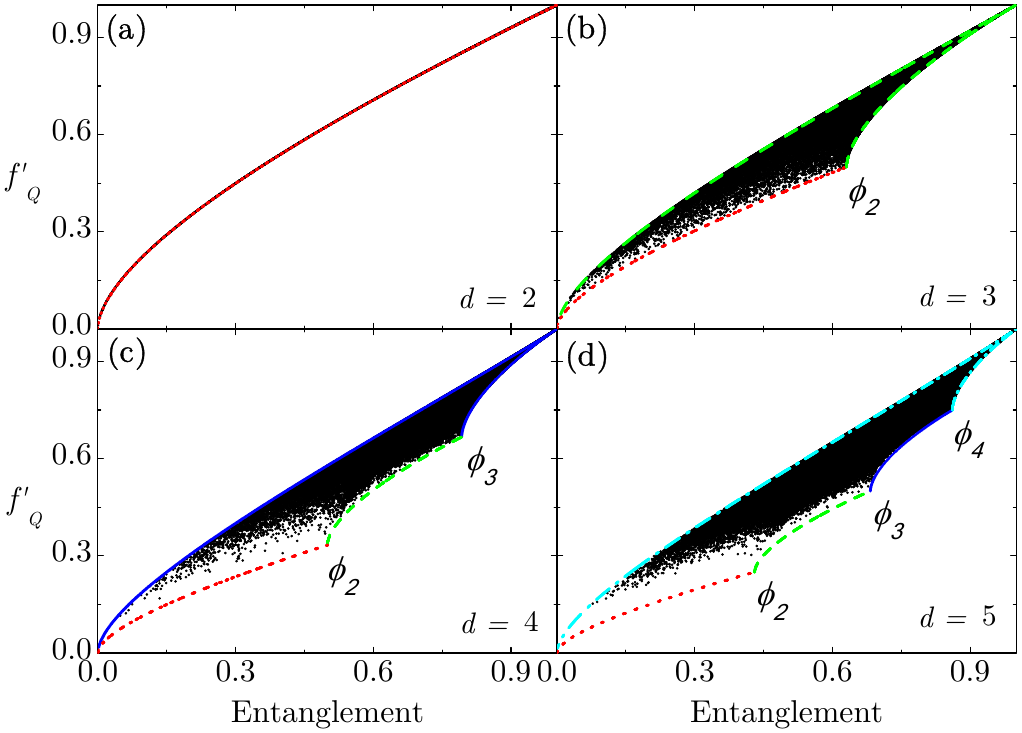}
	\caption{Normalized quantum contribution to the fidelity $f'_Q[=(d+1)f_Q/(d-1)]$ in function of the amount of entanglement in the channel (normalized to $\log d$), for a maximally entangled measurement basis and a set of $N$ random pure states of two entangled qudits, for several values of $d$. \textbf{(a)} $d=2$, $N=10^4$. \textbf{(b)} $d=3$, $N=10^5$. \textbf{(c)} $d=4$, $N=10^5$. \textbf{(d)} $d=5$, $N=10^6$. Each line corresponds to a particular family of ``boundary" states $\varphi_{\mu}$, given by Eq. \ref{Bounds}: \textit{red dotted}: $\mu=1$, \textit{green dashed}: $\mu=2$, \textit{blue solid}: $\mu=3$ and  \textit{cyan dash-dotted}: $\mu=4$. Note also that any intersection point between two family lines corresponds to a maximally symmetric $\nu$-rank state $\phi_{\nu}$ (Eq. \ref{Rank}).}
	\label{Ent_1}
\end{figure}

In order to have a qualitative picture of how the amount of entanglement in the channel is related to the quantum contribution to the fidelity of teleportation $f_Q=\frac{2}{d(d+1)}\sum_{\substack{m,n=0\\j>k=0}}^{d-1} \Re\left(\omega_d^{m(k-j)}\beta_{jm}\beta_{km}^*\gamma_{k\oplus n}\gamma_{j\oplus n}^*\right)$, we produced a sample of random entangled states uniformly distributed in the space of the Schmidt basis \footnote{We produced random states uniformly distributed in the basis $\{\ket{00},\ket{11},\cdots\ket{d-1,d-1}\}$. Note that this parametrization does not cover uniformly the space of pure states for two-qudits. Nevertheless for our purposes it is enough.}. Results of the quantum contribution to the fidelity of teleportation (normalised to one), $f'_Q[=(d+1)f_Q/(d-1)]$, as a function of the amount of entanglement in the channel for the case in which the measurements basis is maximally entangled are presented in figure \ref{Ent_1}. Note that $d=2$ is the only case in which there is a one to one relation between $f'_Q$ and the amount entanglement. In contrast, it is only possible to infer bounds in the extent of channel's entanglement for a given value of the fidelity within the high dimensional case. It is not difficult to show that for each dimensionality $d$, such a boundary is fully described by $d-1$ families of states $\varphi_{\mu}$:
\begin{equation}
\ket{\varphi_{\mu}}=a_{\mu}\ket{00}+\sqrt{\frac{1-a_{\mu}^2}{\mu}}\sum_{k=1}^{\mu}\ket{kk},
\label{Bounds}
\end{equation}
where $\mu=1,\cdots,d-1$, with $a_{\mu}\in\left[0,1/\sqrt{\mu+1}\right]$ for $\mu=1,\cdots,d-2$ and $a_{\mu}\in\left[0,1\right]$ for $\mu=d-1$. It is possible to see that maximally symmetric $\nu$-rank states, 
\begin{equation}
\ket{\phi_{\nu}}=\frac{1}{\sqrt{\nu}}\sum_{k=0}^{\nu-1}\ket{kk},
\label{Rank}
\end{equation}
with $\nu=1,\cdots,d-1$, are extremal cases of the families $\varphi_{\mu}$.

It is worth to remark that the relation between fidelity of teleportation and bounds of entanglement presented above holds whenever the source is capable of providing entangled pairs in pure states. Nevertheless, the previous analysis does not hold when the channel is prepared in a mixed state, given that as shown in the next section, the classical contribution is also modified in this case.
In this respect, by employing the recently introduced concept of \textit{nonclassical teleportation witnesses} \cite{Cavalcanti17t,Carvacho18}, {\v{S}}upi{\'{c}}, Skrzypczyk and Cavalcanti have found that it is possible to infer lower bounds of channel entanglement from teleportation data only \cite{supic2019}, even for scenarios in which Alice and Bob share pairs prepared in some special families of mixed entangled states not capable of attaining fidelity of teleportation values above its classical limit, also known as bound entangled states \cite{Horodecki1998}.

\section{Noise and Kraus operators}
\label{KO}

In addition to the technical limitations in the preparation of the system and realization of measurements in maximally entangled states, noise is an unavoidable feature of real experiments, for this reason it is very important to establish strategies which lead to the improvement of the final results. In the following sections we explore the influence of protecting one or more qudits from noise on the fidelity of teleportation.

To date there are several methods to study the evolution of the state $\hat{\rho}$, associated to an open quantum system \cite{Nielsen10}. One of the most widely used is the Kraus operators formalism, in which the evolution may be modelled by a trace preserving map $\hat{\rho}\to\hat{\rho}'=\sum_{k}\hat{E}_k\hat{\rho}\hat{E}^{\dagger}_k$, where the $\hat{E}_k$'s are known as Kraus operators and satisfy the completeness relation $\sum_{k}\hat{E}^{\dagger}_k\hat{E}_k=\hat{1}$ \cite{Nielsen10}. Under this approach the $\hat{E}_k$ operators contain information about the effects of the system-environment interaction, without the necessity of deepening into the involved physical processes behind. Let us introduce the relation between high-dimensional generalizations of the most known classes of error in quantum systems and Weyl operators.
\begin{figure}
  \centering
  \includegraphics[scale=0.65]{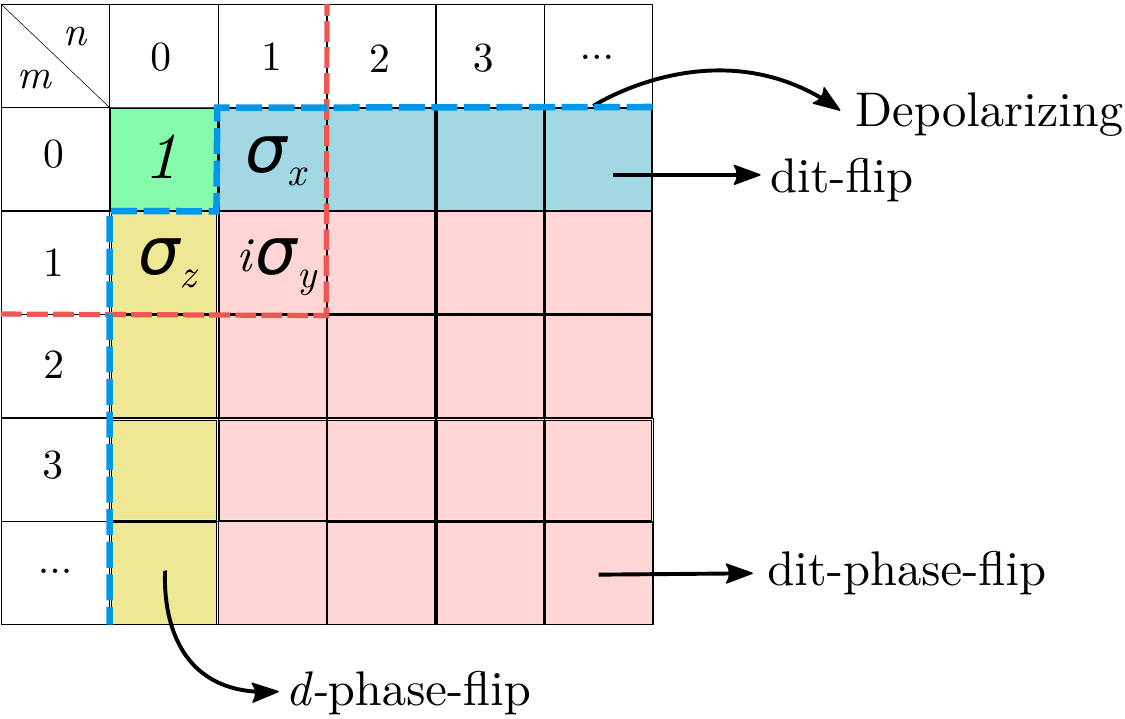}
  \caption{Weyl operators $\hat{U}_{mn}$ and their relation with Kraus operators for several kinds of noise on $d$-dimensional systems. The blue row represents dit-flip like operators, the yellow column $d$-phase-flip like operators and the pink squares are related to matrices corresponding to dit-phase-flip like noise. Note that the three classes mentioned before are employed to define depolarizing noise (blue dashed line). In addition, the set of Pauli matrices corresponds to the nontrivial operators for $d=2$ (red dashed line).}
  \label{Weyl}
\end{figure}
\subsection{Noise and Weyl operators}
%

Kraus operators corresponding to some of the most known instances of quantum noise, namely: bit-flip, phase-flip, bit-phase-flip and depolarizing for $d=2$ and $d=3$ are presented in \cite{Nielsen10} and \cite{Ramzan13} respectively. In addition, it is not difficult to see that arbitrary $d$-dimensional generalizations are proportional to families of Weyl operators $\hat{U}_{mn}$ (Eq. \ref{UKOp}), such a correspondence is illustrated in figure \ref{Weyl}. This class of operators represent a very important tool when dealing with quantum information tasks based on high dimensional systems. Among them it is possible to highlight the quantum teleportation protocol \cite{Bennett93,Weinar13}, representation of quantum states \cite{Bertlmann08}, random unitary evolutions \cite{Chruscinski2015}, quantum computing \cite{Imany2019,Gokhale2019} and quantum error correction \cite{Miller2018}. In the latter case, the elements of set the of Weyl operators (also known as \textit{single-qudit Pauli group}) are often found in literature as products of the form $\omega_d^k\hat{X}^m\hat{Z}^n$, where $\hat{X}$ and $\hat{Z}$ correspond to the operators $\hat{U}_{01}$ and $\hat{U}_{10}$, respectively and $0\leq k,m,n\leq d-1$ \cite{Gokhale2019,Miller2018,Gottesman1999}. In addition, note that the set $\{\hat{U}_{mn}\}$ constitute a natural basis for the $d\times d$ Hilbert-Schmidt space and that Weyl operators are proportional to the set of Pauli matrices for $d=2$ (see Fig. \ref{Weyl}).

Let us briefly describe of each particular noise and the corresponding expressions for the Kraus operators used in this work:

\begin{itemize}
 \item {\textit{dit-flip noise:}} In analogy to bit-flip for $d=2$, this kind of noise considers perturbations that flip $\ket{j}$ either to the state $\ket{j\oplus1}$, $\ket{j\oplus2}$, $~\dots~$, or $\ket{j\oplus d-1}$, with probability $p$. The associated Kraus operators are: $\hat{E}_{00}=\sqrt{1-p}~\hat{U}_{00}$, $\hat{E}_{01}=\sqrt{\frac{p}{d-1}}~\hat{U}_{01}$,$~\dots~$,$~\hat{E}_{0,d-1}=\sqrt{\frac{p}{d-1}}~\hat{U}_{0,d-1}$.

\item {\textit{d-phase-flip noise:}} A qudit $\ket{j}$ subjected to $d$-phase-flip noise may with probability $p$ suffer one out of $d-1$ phase shifts of the form: $\omega_d\ket{j}$, $\omega_d^2\ket{j}$,$~\dots~$,$~\omega_d^{d-1}\ket{j}$. The corresponding Kraus operators are given by: $\hat{E}_{00}=\sqrt{1-p}~\hat{U}_{00}$, $\hat{E}_{10}=\sqrt{\frac{p}{d-1}}~\hat{U}_{10}$,$~\dots~$,$~\hat{E}_{d-1,0}=\sqrt{\frac{p}{d-1}}~\hat{U}_{d-1,0}$.

\item {\textit{dit-phase-flip noise:}} This is a special case in which a combination of both former kinds of noise may take place, e.g. a qudit suffer a flip and a phase shift at the same time. The related Kraus operators are: $\hat{E}_{00}=\sqrt{1-p}~\hat{U}_{00}$ and $\hat{E}_{mn}=\frac{\sqrt{p}}{d-1}~\hat{U}_{mn}$, with $1\le m,n \le d-1$.

\item {\textit{Depolarizing noise:}} Under this, a system initially prepared in an arbitrary state evolves to a maximally mixed state, $\hat{1}/d$ with probability $p$. The Kraus operators for this are given by \cite{Imany2019,Gokhale2019}: $\hat{E}_{00}=\sqrt{1-\frac{d^2-1}{d^2}p}~\hat{U}_{00}$ and $\hat{E}_{mn}=\frac{\sqrt{p}}{d}~\hat{U}_{mn}$, with $0\le m,n \le d-1$,  for $(m,n)\neq(0,0)$. 
\end{itemize}

It is important to mention that the present work does not refer to any particular implementation, for this reason the noise models we have just described consider that possible errors can occur with equal probability. For instance, a qutrit under interactions with the environment leading to a dit-flip noise may jump either one or two levels above, each with probability $p/2$.

The Kraus operators for the classes of noise mentioned above may be written as: $\hat{E}_{mn}=a_{mn}\hat{U}_{mn}$, with coefficients $a_{mn}\in \mathbb{R}$, satisfying $\sum_{mn}a_{mn}^2=1$. Given an arbitrary system initially prepared in a state $\hat{\rho}=\sum_{\vec{k}\vec{l}}\rho_{\vec{k}\vec{l}} ~|\vec{k}\rangle \langle \vec{l}|$,
where $N$ is the number of subsystems, $\vec{k}=(k_1,...,k_N)$ and $0\leq k_j \leq d-1$, the action of a set of Kraus operators 
$\hat{E}_{\vec{k}\vec{l}}=\hat{E}_{k_1l_1}\otimes ... \otimes \hat{E}_{k_Nl_N}=\prod_{j=1}^N a_{k_jl_j}\hat{U}_{k_1l_1}\otimes ... \otimes \hat{U}_{k_Nl_N}$,
transforms $\hat{\rho}$ into $\hat{\rho}'=\sum_{\vec{m}\vec{n}}\rho'_{\vec{m}\vec{n}} ~\dyad{\vec{m}}{\vec{n}}$, with $\rho'_{\vec{m}\vec{n}}$ given by
\begin{equation}
\label{rop}
\rho'_{\vec{m}\vec{n}}=\sum_{\vec{k}\vec{l}}\omega_d^{\vec{k}\cdot(\vec{m}-\vec{n})} \rho_{\vec{m}\oplus\vec{l},\vec{n}\oplus\vec{l}}\prod_{j=1}^N a_{k_jl_j}^2.
\end{equation}
This expression contains the information about noise acting on the whole system and it is very useful in the calculation of a general expression for the fidelity of teleportation (see App. \ref{GEF}).

An important class of noise related to losses in quantum systems is known as \textit{amplitude damping noise}. However, in contrast to the previous cases, the related Kraus operators are not proportional to single Weyl operators. For this reason, calculations of fidelity involving amplitude damping noise were carried out using the standard computational basis. Now let us briefly present its corresponding Kraus operators and the general expression for modified density operator coefficients.

\subsection{Amplitude damping noise}
Amplitude damping noise has been used to model a large amount of phenomena including idle errors in quantum computing \cite{Gokhale2019}, energy dissipation, spontaneous photon emission, attenuation, among others in two level systems \cite{Nielsen10}. A $d$-dimensional generalization was recently introduced in \cite{Dutta16}, the corresponding Kraus operators read
\begin{equation}
\hat{E}_{0}=\dyad{0}{0}+\sqrt{1-p}\sum_{j=1}^{d-1}\dyad{j}{j},
\end{equation}
and
\begin{equation} 
\hat{E}_{j}=\sqrt{p}\dyad{0}{j},
\end{equation}
with $j=1,\cdots,d-1$. This kind of noise may be interpreted in the following way: A $d$-level system interacting with its environment may with probability $p$ lose population from the excited levels, leading the system to the ground state $\ket{0}$.

In general terms, for a system composed by one part only, we can write any Kraus operator as: $\hat{E}_k=\sum_{mn}a_{mn}^{(k)}\dyad{m}{n}$, with coefficients $a_{nm}^{(k)}\in \mathbb{C}$, satisfying $\sum_{kn} a_{mn}^{(k)}a_{ln}^{(k)*}=\delta_{ml}$ (due to the completeness relation). The $N$-party case is a straightforward generalization $\hat{\rho}\to \hat{\rho}'=\sum_{\vec{k}}\hat{E}_{\vec{k}}~\hat{\rho}~\hat{E}_{\vec{k}}^{\dagger}$,
where $\vec{k}=\{k_1,\dots,k_N\}$, as previously and Kraus operators given by $\hat{E}_{\vec{k}}= \hat{E}_{k_1}^{(a_1)}\otimes\hat{E}_{k_2}^{(a_2)}\cdots\otimes\hat{E}_{k_N}^{(a_N)}=\sum_{\vec{m}\vec{n}} \left(\prod_{j=1}^N a_{m_jn_j}^{(k_j)}\right)\dyad{\vec{m}}{\vec{n}}$. In this case the modified density operator coefficients $\rho'_{\vec{p}\vec{q}}$ are reduced to
\begin{equation}
\label{rop2}
\rho'_{\vec{p}\vec{q}}=\sum_{\vec{k}\vec{m}\vec{n}}\left(\prod_{j=1}^N a_{p_jm_j}^{(k_j)}a_{q_jn_j}^{(k_j)*}\right)\rho_{\vec{m}\vec{n}}.
\end{equation}
The coefficients $a_{mn}^{(k_j)}$ may be easily calculated for each particular case as usual: $a_{mn}^{(k_j)}=\bra{m}\hat{E}_{k_j} \ket{n}$.

We have performed calculations of fidelity of teleportation for the cases in which the three qudits involved in the process may be affected by arbitrary combinations of the classes of noise described previously. For details, we refer the reader to Appendix \ref{GEF}. The following sections are devoted to present results concerning some particular cases.

\section{Noise acting on a single qudit}
\label{single}

\subsection{Weyl-like noises}
In this part we consider the case of two qudits fully protected from noise, e.g. an experiment in which the production of pairs of entangled qudits 
is carried out in Alice's location and the Bob's qudit is affected by interacting with the environment during the transportation process.

By direct substitution into the expression for fidelity (Eq. \ref{F3}), it is easy to see that when noise is acting on one qudit only, the fidelity does not depend on the qudit affected. Then, giving continuity to the example given above, we consider that the affected qudit is that on Bob's location. In this case the general expression for the fidelity of teleportation is reduced to:
\begin{equation*}
\expval{F}=\frac{1}{d+1}\Big[1+dc_p^2+c_0^2-c_p^2+(d+1)(c_0^2-c_p^2)f_Q \Big],
\end{equation*}  
which does not depend on the coefficient $c_f$ and for this reason the fidelity corresponding to dit-flip $\expval{F_F}$ and dit-phase-flip $\expval{F_{FP}}$ noises are both equal to:
\begin{equation}
\expval{F_F}=\expval{F_{FP}}=\frac{2}{d+1}\left(1-\frac{p}{2}\right)+f_Q(1-p).
\end{equation}
For $d$-phase-flip noise, the fidelity is reduced to:
\begin{equation}
\label{Fp}
\expval{F_P}=\frac{2}{d+1}+f_Q\left(1-\frac{d}{d-1}p\right),
\end{equation}
the classical fidelity is not affected because phase shifts are exclusive elements of quantum systems. This feature will be explored in more detail in the following subsection.

The corresponding fidelity for depolarizing noise is:
\begin{equation}
\expval{F_{D}}=\frac{2}{d+1}\left(1-\frac{d-1}{2d}p\right)+f_Q\left(1-p\right).
\end{equation}

Note that when we have a maximally entangled channel and measurements, the fidelities $\expval{F_{F}}$, $\expval{F_{P}}$, and $\expval{F_{FP}}$ are all equal to:
\begin{equation}
\expval{F}=1-\frac{d}{d+1}p,
\end{equation}
and the corresponding to depolarizing reduces to:
\begin{equation}
\expval{F_{D}}=1-\frac{d-1}{d}p.
\end{equation}
The noise thresholds [critical noise fractions above which the fidelity of teleportation acquire values below the classical limit, $f_C=2/(d+1)$] are given by $p^*=(d-1)/d$ and $p^*_D=d/(d+1)$.

\subsection{Amplitude-Damping noise}

In this case it is not possible to write a closed expression in terms of the quantum contribution to the fidelity of teleportation $f_Q$, in the same way as in the previous section, for arbitrary dimension, $d$.
For a maximally entangled measurement basis and channel, the fidelity reads
\begin{equation*}
\expval{F_{AD}}=\frac{2}{d+1}\left[\frac{d^2-d+2}{2d}-\frac{(d-1)^2}{2d}p+\frac{d-1}{d}\sqrt{1-p}\right].
\end{equation*} 
The noise threshold $p^*_{AD}$ in this case is given by
\begin{equation*}
p^*_{AD}=\frac{d+2\sqrt{d}}{\left(\sqrt{d}+1\right)^2}.
\end{equation*}
Note that the above expression for the fidelity of teleportation is valid whenever two parts are fully protected and only one qudit is being affected by amplitude damping noise, no matter which one. Nevertheless, this symmetry ceases to appear either when entanglement is not maximal and/or another qudit suffers the action of any kind of noise.


\subsection{Optimization of fidelity under $d$-phase-flip noise in one qudit}
Besides the fact that the classical fidelity is not affected by the presence of $d$-phase-flip noise acting on a single qudit, it is possible to find some other interesting features. By analysing the expression for fidelity (Eq. \ref{Fp}), we see that above a noise threshold $p^*=(d-1)/d$, the coefficient accompanying the quantum contribution becomes negative. This situation may be overcome if we make a phase addition in the measurement basis, as pointed out in \cite{Fortes15} for $d=2$. Without loss of generality and in order to simplify calculations, assume a channel initially prepared in a maximally entangled state and a maximally entangled measurement basis with arbitrary phases $\phi_j$: $\beta_{jm}={\rm e}^{i\phi_j}/\sqrt{d}$, with $\phi_0=0$. The fidelity then holds:
\begin{widetext}
\begin{eqnarray}
\expval{F_P}=\frac{2}{d+1}\left\{1+\frac{1}{d}\left(1-\frac{pd}{d-1}\right)\left( \sum_{k=1}^{d-1}\cos\phi_k+\sum_{k>l=1}^{d-1}\cos(\phi_l-\phi_k)\right)\right\}.
\end{eqnarray}
\end{widetext}

\begin{figure}
	\centering
	\includegraphics[scale=0.325]{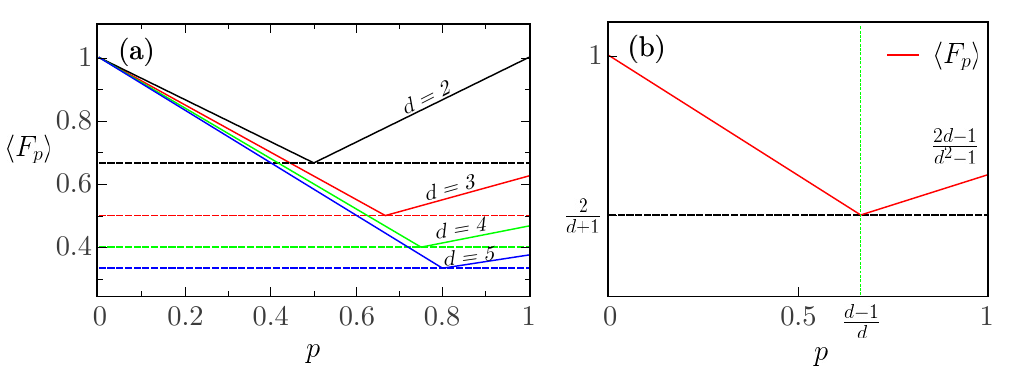}
	\caption{{\it{(solid):}} Optimal fidelity of teleportation for the case in which only one of the qudits may suffer $d$-phase-flip noise.		{\it{(dashed):}} Classical fidelity. {\bf{(a)}} Calculations of $\expval{F_p}$ for $2\leq d\leq5$. {\bf{(b)}} Optimal fidelity of teleportation for arbitrary 
		dimension $d$.}
	\label{Fpab}
\end{figure}

The problem is thus reduced to a optimization procedure in which we search for extremal values (maximum when $p<p^*$ and minimum for $p>p^*$) of the quantum contribution to the fidelity. We carried out analytical calculations up to $d=3$, obtaining following results: For noise fractions below the treshold $p^*$, the whole set of phases are null, as expected. For $p>p^*$, we got $\phi_1=\pi$ for $d=2$ and  $(\phi_1,\phi_2)=(2\pi/3,4\pi/3)$ for $d=3$. The resulting fidelities are plotted in figure \ref{Fpab}(a). Further numerical calculations were performed from which we were able to infer the following expressions for the optimized fidelity: 
\begin{equation}
\expval{F_P}=\begin{cases}
  1-\frac{dp}{d+1} & \for p<p^*\\\\
  \frac{dp+d-1}{d^2-1}& \for p>p^*.
  \end{cases}
\end{equation}

In conclusion, if somehow either Alice or Bob are capable of estimating the associated $d$-phase-flip noise fraction on the affected qudit, then she can improve the fidelity of the teleported state by choosing one out of two measurement basis. The results are summarized in Fig. \ref{Fpab}(b) for arbitrary $d$. As it can be seen, the best improvement is attained by systems composed by qubits. As Fortes and Rigolin have shown \cite{Fortes15}, this feature can be exploited if we permit a part of the system to be strongly affected by phase-flip noise. Nevertheless it is worth to remark that such a recovery in the fidelity of qubit teleportation reported in \cite{Fortes15} may be explained by the fact that an increase in the noise fraction $p$ leads to an effective suppression of the mixedness in the final state as $p\to 1$. Unfortunately this is not the case for arbitrary dimension, given that there are more than one phases involved, thus such a recovery becomes lower as we increase $d$, as it can be seen in Fig. \ref{Fpab}(a).

\section{Noise in more than one qudit}

In this section we explore the case in which protection may be applied in at most one of the qudits \footnote{Here we refer to a \textit{protected qudit} as one whose probability of being affected by any kind of noise is negligible in comparison with that corresponding to other components in the system.}. In order to have a best insight from the results, we assume maximal entanglement in the channel and measurements.  

Before examining several cases in detail let summarize some general results: When entanglement is maximal, given either one, two or three classes of Weyl-like noises acting on the system, the fidelity does not depend on how those are distributed on the qudits. In this way we have $\expval{F_{X,Y,Z}}(p_X,p_Y,p_Z)=\expval{F_{Y,X,Z}}(p_Y,p_X,p_Z)= \expval{F_{Z,X,Y}}(p_Z,p_X,p_Y)=\dots$, where $\expval{F_{X,Y,Z}}(p_X,p_Y,p_Z)$ indicates the teleportation fidelity given that $X$, $Y$ and $Z$ noises are acting on the input, Alice's and Bob's part of the channel, with noise fractions $p_X$, $p_Y$ and $p_Z$ respectively. For instance, any situation in which two qudits may be affected is equivalent to that of having the input protected only, i.e. $\expval{F_{X,\varnothing ,Y}}=\expval{F_{X,Y,\varnothing}}=\expval{F_{\varnothing ,X,Y}}$, where the symbol $``\varnothing"$ stands for a noise-free qudit.

In addition, some interesting results arise when we consider dit-flip, $d$-phase-flip and dit-phase-flip noises: $\expval{F_{F,F,X}}=\expval{F_{P,P,X}}$ with $X=\{\varnothing,FP,D\}$, $\expval{F_{X',Y',F}}=\expval{F_{X',Y',P}}=\expval{F_{X',Y',FP}}$ for $X'\neq Y'=\{\varnothing,D\}$ and  $\expval{F_{X,Y,\varnothing}}=\expval{F_{X,Z,\varnothing}}$ for $X\neq Y\neq Z=\{F,P,FP\}$ explicit expressions are not presented here, however all may be obtained by direct substitution in equation \ref{FG}. 

\begin{figure}
	\centering
	\includegraphics[scale=0.275]{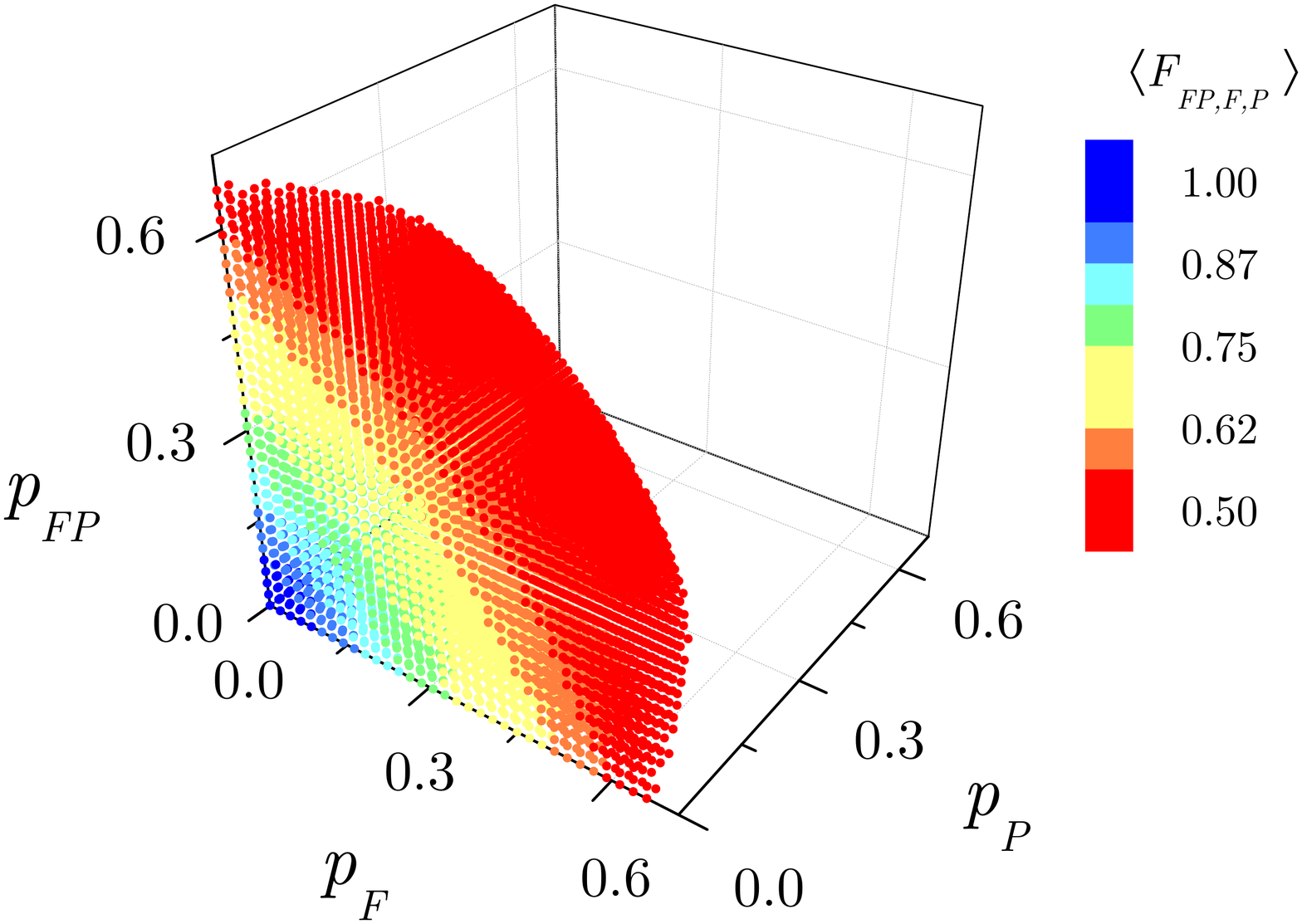}
	\caption{Fidelity of teleportation under a scenario in which the qudits are affected by dit-phase-flip, dit-flip and $d$-phase-flip noises respectively $(FP,F,P)$, and $d=3$, for configurations leading to values above the classical limit, $\expval{F}>f_C=1/2$. It is possible to notice that the larger the noise fraction, the lower the fidelity of teleportation, as expected.}
	\label{Ball}
\end{figure}

With the exception of some specific cases listed below, we observed that whenever the system is subjected to the instances of noise considered in this work, the fidelity of teleportation exhibits quite the same behaviour: the larger the amount of noise, the lower the values attained by $\expval{F}$. As an illustration, we have plotted results of teleportation fidelity in function of the corresponding noise fractions for the scenario $(FP,F,P)$ and $d=3$ in Figure \ref{Ball}. 
Furthermore it was observed that as the dimension increases, the set of noise configurations leading to a fidelity of teleportation above the classical limit becomes larger. Nevertheless this is something we expected, given that in most of the cases despite the teleportation fidelity shows a sharper fall as $d$ increases, the value of the classical limit decreases even more abruptly as shown in Fig. \ref{Fpab}(a).

%

Although in principle it is expected that the fidelity of teleportation tends to decrease for high noise fractions, Fortes and Rigolin \cite{Fortes15} found a set of scenarios in which it is not the case, even with no change in the measurement basis it is possible to get fidelities above the classical limit for high noise fractions. The following part is devoted to explore these cases in detail.

\subsection{\textit{Fighting noise with noise} in qudit teleportation}

\begin{figure}
	\centering
	\includegraphics[scale=0.275]{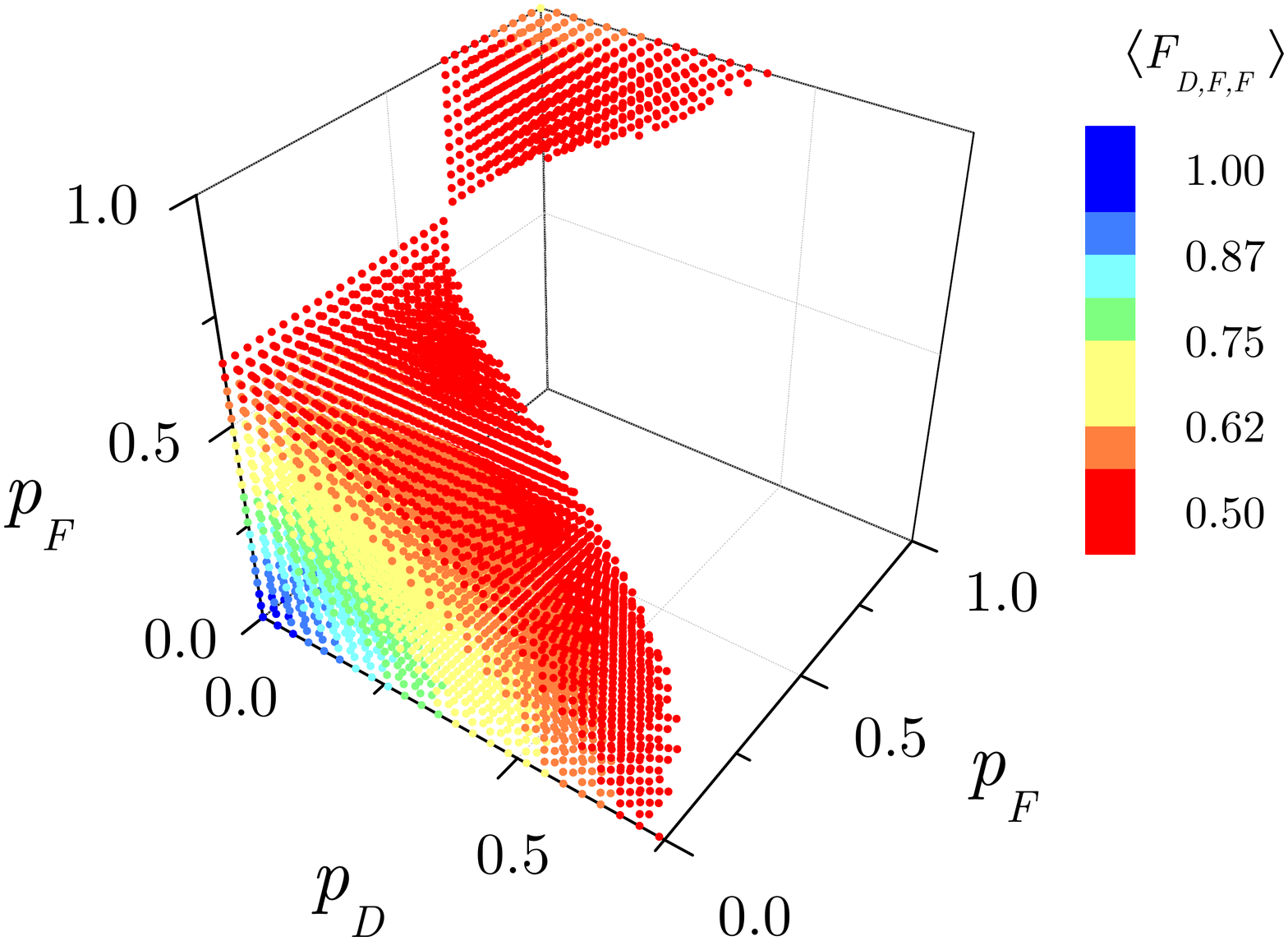}
	\caption{Fidelity of teleportation for the scenario of one  qutrit affected by depolarizing and the other two by trit-flip-noise, for configurations leading to values above the classical limit. Note that there is a region in which the fidelity attains values above $f_C$, even for large noise fractions. See table \ref{Table} for all possible noise configurations presenting this behaviour.}
	\label{DFF}
\end{figure}

We have performed an exhaustive search of scenarios in which the addition of noise to the system leads to an enhancement in the execution of the protocol without carrying out a basis change, obtaining the following extremal scenarios: $(\varnothing,F,F)$, $(\varnothing,P,P)$, [$(\varnothing,FP,FP)$, for $d=2$ only] and commutations. Under these, whenever both associated noise fractions get values either lower or higher than the noise threshold $p^*=1-1/d$, the fidelity attains values above the classical limit, with $\expval{F}\to (2d-1)/(d^2-1)$ as the noise fractions approach to 1. Notice though that given the dependency of the classical fidelity on $d$ $[f_C\sim \mathcal{O}(d^{-1})]$, thus the larger the dimensionality, the more negligible becomes the gap between $\expval{F}$ and $f_C$ in the regime of high noise fractions. Moreover, note that a perfect restoration in the fidelity $(\expval{F}=1)$, is achieved by $d=2$ only. It happens because the error in one of the qubits is globally corrected by the action of the same kind of error in another part of the system, while such a correction can be well succeeded for high dimensional systems only probabilistically, due to the fact that in this case there are several possible final configurations (e.g. a qutrit $\ket{j}$ may be flipped in two different ways: either $\ket{j\oplus 1}$ or $\ket{j\oplus 2}$). Furthermore in the limit $p\to 1$, $d=2$ is the only case in which there is just one Kraus operator acting on the state and in this way the purity of the subsystem is not altered.
\begin{figure}
	\centering
	\includegraphics[scale=0.275]{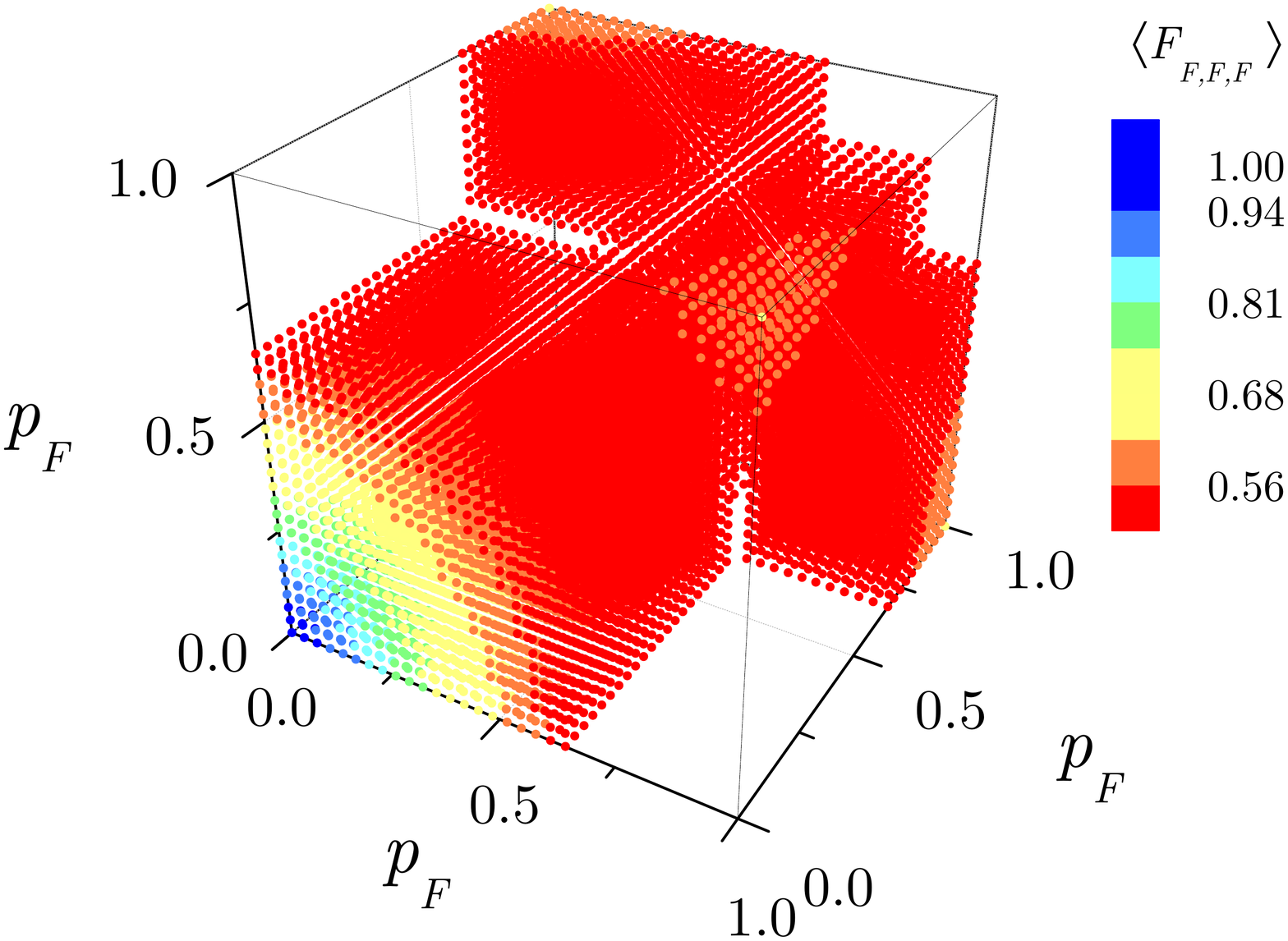}
	\caption{Fidelity of teleportation for the scenario in which all the parts are affected by dit-flip noise, and $d=3$. The axes represent the noise fraction associated to each qudit involved in the teleportation process. Note that fidelity of teleportation values above the classical limit are obtained either if all the noise fractions are lower or two of them are higher than the noise threshold $p^*=1-1/d$.}
	\label{FFF}
\end{figure}
Along with the scenarios we have just described, there are some other instances for which even adding noise to the third qudit, the system is still able to exhibit fidelities beyond its classical value, as exposed in figure \ref{FFF} for a system under the action of dit-flip noise independently on its three components. For this, the teleportation fidelity reaches a value above $f_C$ whenever either the noise fractions are all below or two of them are above a noise threshold $p^*=1-1/d$. It is important to remark that the scenario $(P,P,P)$ exhibits the same behaviour.
The other relevant situations correspond to adding an arbitrary class of noise to the input's qudit and dit-flip or $d$-phase-flip noise on the qudits in the channel. For all these instances the fidelity of teleportation behaves in the same way. In particular, results corresponding to the scenario $(D,F,F)$ are presented in figure \ref{DFF}. Note that there are two separated regions satisfying $\expval{F}>f_C$: the usual corresponding to all noise fractions below the threshold, and on the other hand, that in which both noise fractions associated to dit-flip errors are above the threshold $p^*=1-1/d$, extended throughout the $p_D$ axis up to a maximal depolarizing noise fraction $p_D^*$ equal to $d/(d^2-d+1)$. The whole scenarios are summarized in Table \ref{Table}, as well as the maximal noise fractions on the input's qudit and possible commutations among the involved parts.

\begin{table}[b]
	\caption{\label{Tab} Scenarios in which the fidelity of teleportation $\expval{F_{X,Y,Z}}$ attains values above the classical limit for $p_Y,p_Z>p^*=1-1/d$, and corresponding maximal noise fractions $p_X^*$. The last column indicates the cases where the fidelity does not change upon commutation between affected qudits and noise, i.e.  $\expval{F_{X,Y,Z}}=\expval{F_{Z,Y,X}}=\expval{F_{Y,X,Z}}=\cdots$.}
	\begin{ruledtabular}
		\begin{tabular}{ccccc}
			$X$ & $Y$ & $Z$ & $p_X^*$ &Commutations?\\
			\hline
			$P$ & $F$ & $F$ & $1/d$ & Yes \\
			$FP$ & $F$ & $F$ & $1/d$ & Yes \\			
			$D$ & $F$ & $F$ & $d/(d^2-d+1)$ & Yes \\			
			$AD$ & $F$ & $F$ & $-$\footnotemark[1] & No \\			
			$F$ & $P$ & $P$ & $1/d$ & Yes \\
			$FP$ & $P$ & $P$ & $1/d$ & Yes \\			
			$D$ & $P$ & $P$ & $d/(d^2-d+1)$ & Yes \\			
			$AD$ & $P$ & $P$ & $-$\footnotemark[1] & No \\
		\end{tabular}
	\end{ruledtabular}
	\footnotetext[1]{$p^*=2\sqrt{2}-2$, for $d=2$.}
	\label{Table}
\end{table}

\begin{figure}
	\centering
	\includegraphics[scale=0.275]{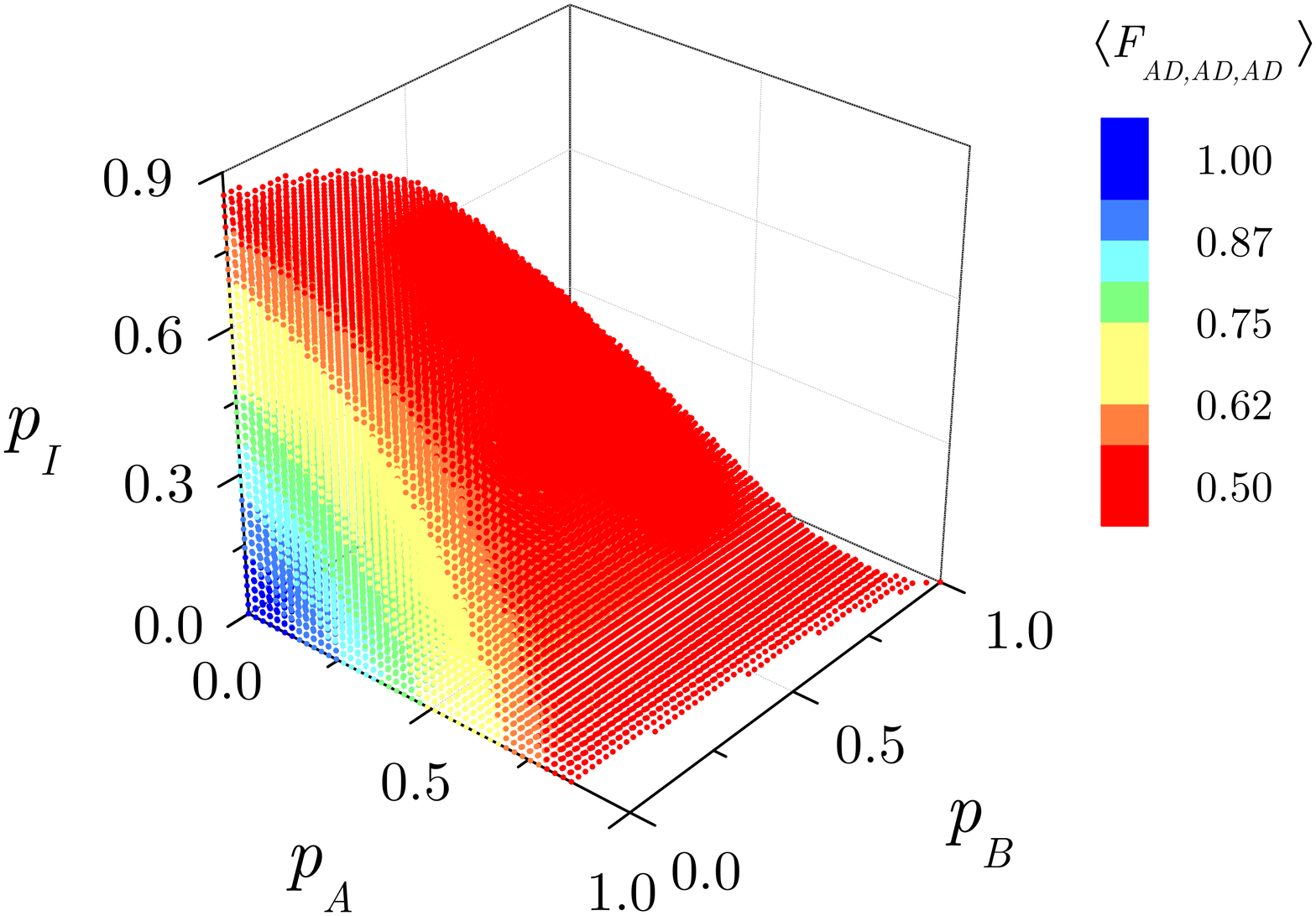}
	\caption{Fidelity of teleportation for a scenario in which all parts are affected by amplitude damping noise $(AD,AD,AD)$ and $d= 3$, in function of the noise fractions associated to each qudit in the channel. Note that in the limit $p_I\to0$, only a small region of noise parameters lead to fidelities below its classical value, $f_C=1/2$. Furthermore, as the amount of noise in the channel increases, the fidelity approaches to its classical limit.}
	\label{AAA}
\end{figure}

Another interesting case observed in \cite{Fortes15} (also in \cite{Knoll14}, using the singlet fraction instead \cite{Horodecki1999}), is the scenario $(\varnothing,AD,AD)$, for $d=2$. In this, the fidelity of teleportation has an asymptotic tendency to the classical limit when the noise fractions approach to their maximal value, as it can be observed in Fig. \ref{AAA} for $d=3$. Another feature we were able to infer is that the regions of noise parameters corresponding to $\expval{F}<f_C$ are relatively small when compared to other scenarios involving noise on two parts. As an illustration, for $d=2$ it covers $\sim24,44\%$ out of the whole configurations and exhibits a tendency to decrease with $d$, for instance it is only about $15,4\%$ for $d=5$.

\section{Conclusion}

We have carried out a characterization of the qudit teleportation protocol in realistic scenarios, generalizing the results recently obtained for the case of qubits \cite{Fortes15}. Under this approach, errors may be introduced to the system either by imperfections in the preparation and measurements and/or by the unavoidable interaction with the environment. We have been able to establish relations among the fidelity of teleportation and the amount of entanglement in the channel for the noiseless scenario. We performed an exhaustive search of all possible kinds of behaviour in the fidelity of teleportation, finding four predominant sets: \textit{i)} The most typical and intuitive may be described as a decay in the fidelity with the noise fractions, depicted in figure \ref{Ball}. \textit{ii)} The scenario $(F,F,F)$ or $(P,P,P)$ (Figure \ref{FFF}), in which the presence of noise in one qudit leads to an effective partial correction of the same kind of error on a different part. \textit{iii)} An intermediate behaviour between the previous scenarios, on the one hand a decay in the fidelity and on the other one a partial error correction, summarized in Table \ref{Table} and Figure \ref{DFF}. And \textit{iv)} the corresponding to any kind of noise in the input and Amplitude damping in the channel, depicted in Figure \ref{AAA}. 
Furthermore it was possible to note that it is possible to partially correct errors by a basis change, for a very specific case ($P,\varnothing,\varnothing$). 
It is left as an open question whether considering special scenarios in which the nature of the interactions favours some sub-classes of noise, can lead to an improvement of the performance of the protocol of high dimensional quantum teleportation.
As a final remark, it would be very interesting to find further strategies leading to the improvement of the protocol of high-dimensional quantum teleportation, such as the presence of non-local memory effects \cite{Laine2014}. 

\appendix

\section{Parametrization for pure qudit states}	
\label{ApA}
Any arbitrary pure state of a qudit $\ket{\phi}=\sum_{j=0}^{d-1}\alpha_j\ket{j}$ may be parametrized as:
\begin{equation*}
\label{Par}
\alpha_j=\begin{cases}
  \cos\theta_0 & \for j=0\\
  \sin\theta_0~\dots~\sin\theta_{j-1}\cos\theta_{j}~{\rm e}^{i\phi_j}& \for 1\leq j\leq d-2\\
  \sin\theta_0~\dots~\sin\theta_{d-2}~{\rm e}^{i\phi_{d-1}}& \for j=d-1,
  \end{cases}
\end{equation*}
with $0<\theta_j\leq \pi/2$ and $0<\phi_j\leq 2\pi$.

Under this parametrization the invariant volume element $\dd\Gamma_d$ is given by \cite{Caves01,Bengtsson07,Zyczkowski01}:
\begin{eqnarray*}
\dd\Gamma_d=\sin^{2d-3}\theta_{0}&&\dots \sin\theta_{d-2}\cos\theta_0\dots\\
&&\dots\cos\theta_{d-2}\dd\theta_0\dots\dd\theta_{d-2}\dd\phi_1\dots\dd\phi_{d-1},
\end{eqnarray*}
in a compact form:
\begin{equation}
\label{dG}
\dd\Gamma_d=\prod_{j=0}^{d-2}\sin^{2d-2j-3}\theta_{j}\cos\theta_j~\dd\theta_j~\dd\phi_{j+1}.
\end{equation}
The total volume $V_d=\int \dd\Gamma_d$ may be easily calculated and is equal to:
\begin{equation}
V_d=\frac{\pi^{d-1}}{(d-1)!}.
\end{equation}
\section{Calculation of $\expval{\alpha_{j}\alpha_{k}^*\alpha_{l}\alpha_{m}^*}$}
\label{ApC}

The calculation of $\expval{\alpha_{j}\alpha_{k}^*\alpha_{l}\alpha_{m}^*}=\frac{1}{V_d}\int \dd \Gamma_d~\alpha_{j}\alpha_{k}^*\alpha_{l}\alpha_{m}^*$ may be facilitated if we take into account some symmetries. First of all, note that the volume element $\dd \Gamma_d$ does not depend explicitly on the phases $\phi_{j}$. Moreover, the state coefficients $\alpha_{j}$ are proportional to $\exp(i\phi_{j})$, then the only way in which the integration does not vanish is having both: the coefficient and its conjugate inside the argument in order to cancel the corresponding phases. In this way we must have: $\expval{\alpha_{j}\alpha_{k}^*\alpha_{l}\alpha_{m}^*}\propto\big(\delta_{jk}\delta_{lm}+\delta_{jm}\delta_{kl}\big)$. Let us determine the proportionality constant. For simplicity we only show calculations for $\expval{\abs{\alpha_0}^4}$, nevertheless as the generated states are uniformly distributed, then any choice is equivalent. The integration reads:
\begin{eqnarray*}
\expval{\abs{\alpha_0}^4}=\frac{(d-1)!}{\pi^{d-1}}\int \prod_{j=0}^{d-2}&&\sin^{2d-2j-3}\theta_{j}\times\\
&&\times\cos\theta_j~\dd\theta_j~\dd\phi_{j+1} \cos^4\theta_0.
\end{eqnarray*}
It is not hard to show that integrations of the kind above have the following solutions:
\begin{eqnarray*}
I^n_m=&&\int_{0}^{\pi/2}\sin^mx\cos^{n+1}x~\dd x\\
=&&\sum_{k=0}^{n/2}\frac{(-1)^k\left(\frac{n}{2}\right)!}{k!\left(\frac{n}{2}-k\right)!}\frac{1}{2k+m+1},
\label{Imn}
\end{eqnarray*}
for $n=0,2,4,\cdots$ and $m>0$.
Thus $\expval{\abs{\alpha_0}^4}$ is reduced to
\begin{equation}
\expval{\abs{\alpha_0}^4}=2^{d-1}(d-1)!~I_{2d-3}^{4}~\prod_{j=1}^{d-2}I_{2d-2j-3}^{0}.
\end{equation}
It is straightforward to see that $I_{2d-2j-3}^{0}=\frac{1}{2(d-j-1)}$ and $I_{2d-3}^{4}=\frac{1}{(d+1)d(d-1)}$, then
\begin{equation*}
\expval{\abs{\alpha_0}^4}=2^{d-1}(d-1)!\frac{1}{(d+1)d(d-1)}~\prod_{j=1}^{d-2}\frac{1}{2(d-j-1)}.
\end{equation*}
This expression reduces to
\begin{equation}
\expval{\abs{\alpha_0}^4}=\frac{2}{d(d+1)}.
\end{equation}
Back to the general case, it is possible infer that the proportionality factor must be equal to $\frac{1}{d(d+1)}$. In this way we have
\begin{equation}
\label{Integ}
\expval{\alpha_{j}\alpha_{k}^*\alpha_{l}\alpha_{m}^*}=\frac{1}{d(d+1)}\big(\delta_{jk}\delta_{lm}+\delta_{jm}\delta_{kl}\big).
\end{equation}
The result above is very useful in the calculation of reduced expressions for the average fidelity of teleportation (see next Appendix).

\section{General expressions for fidelity of teleportation}
\label{GEF}

This appendix is devoted to present the derivation of general expressions for the fidelity of teleportation within noisy environments under two approaches. The first contemplates the cases in which the Kraus operators associated to the classes of noise involved are proportional to Weyl operators. In the second part we consider Kraus operators written in the standard computational basis in order to consider the cases in which amplitude damping noise may take place in any part of the system.

\subsection{Weyl-like noises}

The noise coefficient associated to the input qudit $a_{jk}$ may be expressed as a superposition of the contributions of each region in figure \ref{Weyl}: $a_0$, noiseless region (green); $a_f$, flip region (blue); $a_p$, phase flip region (yellow) and $a_c$ for the region of combination of flip and phase-flip (red). In this way, the squared noise coefficient reads
\begin{eqnarray}
\label{NCF}
a_{jk}^2=a_0^2&&\delta_{j,0}\delta_{k,0}+a_f^2\delta_{j,0}\sum_{n=1}^{d-1}\delta_{k,n}+\nonumber \\
&&+a_p^2\delta_{k,0}\sum_{m=1}^{d-1}\delta_{j,m}+ a_c^2\sum_{m,n=1}^{d-1}\delta_{j,m}\delta_{k,n}.
\end{eqnarray}
Thus we have the following correspondences between noise and reduced coefficients: dit-flip: $a_p=a_c=0$, phase-flip: $a_f=a_c=0$, dit-phase-flip: $a_p=a_f=0$ and
$a_f=a_p=a_c$ for depolarizing.

After some steps, the fidelity of teleportation $F=\sum_{mn}\Tr\left\{\dyad{\phi}{\phi}\hat{\rho}_{mn}\right\}$ takes the form:
\begin{equation*}
F=\sum_{\substack{jkmn \\ \mu\nu=0}}^{d-1}\alpha_m\alpha_n^*\beta_{j\mu}\beta_{k\mu}^*\omega_d^{\mu(n-m)}\rho'_{k,k\oplus\nu,n\oplus\nu,j,j\oplus\nu,m\oplus\nu},
\end{equation*}
using equation (\ref{rop}) and assuming a channel initially prepared in a pure state $\ket{\psi}=\sum_{k=0}^{d-1}\gamma_k \ket{kk}$, the fidelity of teleportation holds:
\begin{widetext}
\begin{eqnarray}
F=\sum_{\substack{jk\mu\nu p_1p_2p_3  \\ q_1q_2q_3=0}}^{d-1} \alpha_{j\oplus q_2\ominus q_3} \alpha_{k\oplus q_2\ominus q_3}^* \alpha_{j\oplus q_1}^* \alpha_{k\oplus q_1} \beta_{j\mu}\beta_{k\mu}^* \omega_d^{(k-j)(\mu+p_1+p_2+p_3)} \gamma_{k\oplus \nu \oplus q_2} \gamma_{j\oplus \nu \oplus q_2}^* a_{p_1q_1}^2b_{p_2q_2}^2c_{p_3q_3}^2,
\end{eqnarray}
%
where $b_{p_2q_2}$ and $c_{p_3q_3}$ are the noise coefficients corresponding to the channel qudits respectively. By using the result of Appendix \ref{ApC} (eq. \ref{Integ}) and after calculations the average fidelity takes the form
%
\begin{eqnarray}
\label{F3}
\expval{F}=\frac{1}{d+1}\left\{1+\frac{1}{d}\sum_{\substack{jk\mu\nu p_1p_2p_3  \\ q_1q_2q_3=0}}^{d-1} \beta_{j\mu}\beta_{k\mu}^* \omega_d^{(k-j)(\mu+p_1+p_2+p_3)} \gamma_{k\oplus \nu \oplus q_2}\gamma_{j\oplus \nu \oplus q_2}^* a_{p_1q_1}^2b_{p_2q_2}^2c_{p_3q_3}^2 \delta_{q_2,q_1\oplus q_3}  \vphantom{\sum_{\substack{j\\q}}^d}\right\}.	
\end{eqnarray}
%

Note that for the noiseless case the noise coefficients read $a_{p_jq_j}=\delta_{p_j,0}\delta_{q_j,0}$ (the same for $b_{p_jq_j}$ and $c_{p_jq_j}$), then the fidelity reduces to the equation obtained in \ref{NFE} for a noise free environment, as expected.
By using analogous expressions for the noise coefficients of the channel qudits $b_{jk}$ and $c_{jk}$ (Eq. \ref{NCF}), substituting into equation \ref{F3}, and after some calculations the fidelity of teleportation becomes: 	
%
\begin{eqnarray}
\label{FG}
&\expval{F}=\frac{1}{d+1}\Bigg(1+d\bigg\{b_p^2\Big[a_0^2c_0^2+(d-1)a_f^2c_f^2 \Big]+\Big[b_0^2+(d-2)b_p^2\Big]\Big[a_p^2c_0^2+a_0^2c_p^2+(d-1)\left(a_f^2c_c^2+a_c^2c_f^2 \right) \Big]+ \notag \\
&+\Big[(d-2)b_0^2+(d^2-3d+3)b_p^2\Big]\Big[a_p^2c_p^2+(d-1)a_c^2c_c^2 \Big] \bigg\}+\notag \\
&+d(d-1)\bigg\{\Big[(d-2)b_f^2+(d^2-3d+3)b_c^2\Big] \Big[a_p^2c_c^2+a_c^2c_p^2+(d-2)a_c^2c_c^2\Big]+b_c^2\Big[a_f^2c_0^2+a_0^2c_f^2+(d-2)a_f^2c_f^2\Big]+\notag \\
&+\Big[b_f^2+(d-2)b_c^2\Big]\Big[a_c^2c_0^2+a_0^2c_c^2+a_f^2c_p^2+a_p^2c_f^2+(d-2)\left(a_f^2c_c^2+a_c^2c_f^2 \right)\Big]\bigg\}+ \notag \\
&+\left(b_0^2-b_p^2\right)\Big[\left(a_0^2-a_p^2\right)\left(c_0^2-c_p^2\right)+(d-1)\left(a_f^2-a_c^2\right)\left(c_f^2-c_c^2\right)\Big] \Big(1+(d+1)f_Q \Big)+\notag \\
&+\left(b_f^2-b_c^2\right)\Big[\left(a_0^2-a_p^2\right)\left(c_f^2-c_c^2\right)+\left(a_f^2-a_c^2\right)\left(c_0^2-c_p^2\right)+(d-2)\left(a_f^2-a_c^2\right)\left(c_f^2-c_c^2\right)\Big]\tilde{f}\Bigg),
\end{eqnarray}
where $f_Q$ is the quantum contribution to the fidelity of teleportation in the absence of noise and $\tilde{f}$ is related to the channel and measurement coefficients as:
\begin{equation}
\tilde{f}=\frac{1}{d}\sum_{\substack{jk\mu\nu=0\\q=1}}^{d-1} \beta_{j\mu}\beta_{k\mu}^*\omega_d^{\mu(k-j)}\gamma_{k\oplus \nu\oplus q}\gamma_{j\oplus \nu\oplus q}^*,
\end{equation}
attaining its highest value $d(d-1)$ when the entanglement in the channel and measurements is maximal.
\subsection{Kraus operators in the standard computational basis}

Let us calculate the fidelity of teleportation. Substituting Eq. \ref{rop2} in Eq. \ref{F0}, we have
\begin{equation*}
F=\sum_{\substack{jkmn\mu\nu\\n_1n_2p_1p_2\\k_1k_2k_3}}\alpha_m\alpha_n^*\alpha_{n_1}\alpha_{p_1}^*\beta_{j\mu}\beta_{k\mu}^*\gamma_{n_2}\gamma_{p_2}^*\omega_d^{\mu(n-m)}a_{k,n_1}^{(k_1)}b_{k\oplus\nu,n_2}^{(k_2)}c_{n\oplus\nu,n_2}^{(k_3)}a_{j,p_1}^{(k_1)*}b_{j\oplus\nu,p_2}^{(k_2)*}c_{m\oplus\nu,p_2}^{(k_3)*}.
\end{equation*}
Analogously to the previous treatment, using the results of Appendix \ref{ApC} and after some calculations, the average fidelity of teleportation holds
\begin{multline}
\label{FA2}
\expval{F}=\frac{1}{d(d+1)}\sum_{\substack{jkm\mu\nu\\n_1n_2p_2\\k_1k_2k_3}}\beta_{j\mu}\beta_{k\mu}^*\gamma_{n_2}\gamma_{p_2}^*a_{k,n_1}^{(k_1)} b_{k\oplus\nu,n_2}^{(k_2)}b_{j\oplus\nu,p_2}^{(k_2)*}c_{m\oplus\nu,p_2}^{(k_3)*}\left[a_{j,n_1}^{(k_1)*}c_{m\oplus\nu,n_2}^{(k_3)}+\omega_d^{\mu(n_1-m)} a_{j,m}^{(k_1)*}c_{n_1\oplus\nu,n_2}^{(k_3)}\right].
\end{multline}
\end{widetext}
It is worth to mention that the expression for fidelity above (Eq. \ref{FA2}) reproduces the whole results of Fortes and Rigolin \cite{Fortes15} for the case of qubits ($d=2$).
\begin{acknowledgments}
It is a pleasure to thank M\'arcio M. Cunha for enlightening discussions, valuable remarks and collaboration at early stages of this project. Useful suggestions by Nadja Bernardes, Fernando de Melo, Rafael Chaves, Carlos Batista, Marcelo Terra Cunha and Fernando Parisio are also acknowledged. This work was supported by a postgraduate grant from Conselho Nacional de Desenvolvimento Cient\'ifico e Tecnol\'ogico (CNPq). Support from Instituto Nacional de Ci\^encia e Tecnologia-Informa\c{c}\~ao Qu\^antica (INCT-IQ) is also acknowledged.
\end{acknowledgments}
\bibliography{References}
\end{document}